\documentclass[pre,aps,superscriptaddress,twocolumn,floatfix,amssymb]{revtex4}
\usepackage{epsfig}

\bibpunct[,]{(}{)}{;}{a}{,}{,}

\begin{document}
\title{Cooperation and emergence of role differentiation in the dynamics of social
networks}

\author{V\'{\i}ctor M. Egu\'{\i}luz}
\affiliation{Instituto Mediterr\'aneo de Estudios Avanzados IMEDEA
(CSIC-UIB), E07122 Palma de Mallorca, Spain}
\author{Mart\'{\i}n G. Zimmermann}
\affiliation{Facultad de Ciencias Exactas y Naturales, Universidad
de Buenos Aires, 1428 Buenos Aires, Argentina}
\author{Camilo J. Cela-Conde}
\affiliation{Departamento de Filosof\'{\i}a, Universidad de las
Islas Baleares, E07122 Palma de Mallorca, Spain}
\author{Maxi San Miguel}
\affiliation{Instituto Mediterr\'aneo de Estudios Avanzados IMEDEA
(CSIC-UIB), E07122 Palma de Mallorca, Spain}
\email{maxi@imedea.uib.es}

\date{\today}


\begin{abstract}
Models of cooperation grounded on social networks and on the
ability of individuals to choose actions and partners aim to
describe human social behavior. Extensive computer simulations of
these models give important insight in the social mechanisms
behind emerging collective behavior. We consider the entangled
co-evolution of actions and social structure in a new version of a
spatial Prisoner's Dilemma model that naturally gives way to a
process of social differentiation. Diverse social roles emerge
from the dynamics of the system: leaders -individuals getting a
large payoff who are imitated by a considerable fraction of the
population-, conformists -unsatisfied cooperative agents that keep
cooperating-, and exploiters -Defectors with a payoff larger than
the average one obtained by Cooperators. The dynamics generates a
social network that can have the topology of a small world
network. The network has a strong hierarchical structure in which
the leaders play an essential role sustaining a highly cooperative
stable regime. The social structure is shown to be very sensitive
to perturbations acting on the leaders. Such perturbations produce
social crisis described as dynamical cascades that propagate
through the network.
\end{abstract}
\maketitle

\section{Introduction}

Social traps (Platt 1973) are situations in which rational
individual choices result in an undesirable collective outcome for
the social group. A well known example is the problem of
establishing cooperation in a social group (Axelrod 1994) or, more
generally, problems of public goods (Olson 1965, Hardin 1968). It
is often recognized that the understanding of collective social
behavior, once individual attitudes are known, requires taking
into account the interactions among the individuals of the group,
and that these are mediated by a network of social relations
(Granovetter 1973, 1978). Such network constitutes the social
structure of the group. The embeddedness of the interactions in
the social structure (Granovetter 1985) has been identified as a
main ingredient in explaining the evolution of cooperation (Macy
and Skvoretz 1998). It is also well documented (Lazer 2001), that
in the same way that the actions of the individuals are affected
by the social network, the network is not an exogenous structure,
but it is rather created by individual choices. However, there are
not many specific models of social dynamics that incorporate
explicitly the concept of co-evolution of individual and network
(Lazer 2001). In fact, in the long term research agenda posed by
Macy (1991) a central point is that the structure of the network
should not be considered as given, but as a variable. The question
posed is {\it ...to explore how social structure might evolve in
tandem with the collective action it makes possible}. This goes
further beyond models in which there is some network evolution
decoupled from the evolution of the actions of the individuals in
the group. In the context of reciprocal altruism and the building
of cooperation, this general question was implicitly considered
within a game theory simulation model named the Social Evolution
Model (SEM) \citep{Zeggelink00,deVos01}. In this
paper we address explicitly the problem of co-evolution in a
version of the Prisoner's Dilemma (PD) (Rapoport and Chammah 1965)
in which players interact through a network that adapts to the
results of the game, and therefore to the actions of the players.
We focus on the resulting type of social structure (cohesive group
versus social hierarchy), as well as on the dynamical mechanisms
needed to produce the topological properties of the network of
interactions that stabilize a collective cooperative behavior.

A main result of our analysis, extensively based on computer
simulations (Zimmermann et al. 2001), is the {\it emergence} of a
process of social differentiation together with the building-up of
a network with hierarchical relations. Starting from random
partnership among equivalent individuals, a social structure
emerges. In this emerging structure, the topology of the network
of social relations identifies individual with different social
roles: leaders, conformists and exploiters. These roles have been
spontaneously selected during the complex adaptive evolution of
the social group that entails a learning process. This social
structure sustains global cooperation with exploiters surviving in
hierarchical chains in the network. This result is at variance
with other simulation models, such as SEM, in which partner
selection leads to the formation of cohesive egalitarian clusters
\citep{Zeggelink00,deVos01}: the emergence of these groups in a
socially segmented population, with exclusion of free riders,
seems to be the basis of the survival of cooperation in these
other studies. Concerning the topology of the co-evolved network
that sustains cooperation, we find a hierarchical
network with an exponential connectivity distribution.
Our results show that clustering is not needed to sustain
cooperation. However the additional inclusion of local neighboring
partner selection in the model generates the celebrated
small-world connectivity (Watts 1999) of the network.

From a philosophical point of view, the concept of {\it emergence}
is used in contemporary sociology in contradictory ways, being its
proper meaning debatable (Sawyer 2001). We use it here as it
appears in multiagent models of social systems (Gilbert and Conte
1965; Schelling 1978; Epstein and Axtell 1996; Axelrod 1997). In
this context it refers, as often also used in the natural
sciences, to complex dynamical behavior or properties that cannot
be reduced to, or predicted from a detailed description of the
units that compose a system, so that the reductionist hypothesis
does not imply a constructionist one (Anderson 1972). These ideas
put forward a hierarchical structure of science in which different
concepts and descriptions are needed at different levels. The key
idea is that sociology cannot be reduced to psychology, as
molecular biology is not just applied chemistry (Anderson 1972).
From this perspective, and recognizing the limitations of game
theory in describing human interactions, we learn much from a
metaphor like the Prisoner's Dilemma when considering emergent
properties which are not simply linked to special features of
individual human behavior.

A canonical example of emergence is the V shape of bird flocks
(Sawyer 2001). The shape of the flock and the fact that a
particular bird plays the role of a leader, with other birds
lining up behind it, is a nontrivial result of simple interaction
rules, but the leader is neither genetically determined nor
externally appointed. Our results parallel this example, showing
how equivalent agents confronted with the choice of an action
(cooperating or defecting) and with the possibility of choosing
partners, differentiate and acquire different roles while building
up a social structure. This process is the result of interactions
among neighbors that determine an entangled co-evolution of the
choice of actions and of the social network. A particular
enlightening example of the general process that we address here
is the cooperative relations among researchers in a given
scientific field. A given researcher might choose to work or not
to work together with other scientist and, depending on the degree
of success of this collaboration, he might search in the community
to find other scientists with which a profitable cooperative
relation can be alternatively established. This co-evolution
process results in a network of collaborations in which different
scientists play different roles.

Our results might be compared with other studies of group
formation beyond those associated with the SEM. In the theory of
group stability of Carley (1991), as it also happens in our simple
model of equivalent agents, individuals assume multiple roles.
Also there, these diverse roles produce, at the group level, group
stability or change depending on the social structure. However,
while the theory of Carley is based on sharing of knowledge, this
aspect is missing in our model. Stability is in our case strictly
dependent on the adapting structure of the network of
interactions.

In the remainder of this introduction we give a brief discussion
of the main ingredients of our analysis. These ingredients are
models of cooperation, models of social networks, and the
implications of the will of the agents that manifests itself in
the ability of making choices. The second section of the paper
presents our model and some general conclusions from its analysis.
The main body of our computer simulation results are summarized in
Section 3. We conclude with a discussion of results and
limitations of our study.

\subsection{Routes to cooperation}

Why do people cooperate? Why cooperation is empirically observed
when there is a conflict between the self-interest and the common
good? Classical answers to these questions are formulated in terms
of kin (Hamilton 1964) or group selection (Wilson and Sober 1994),
or cooperation based upon reciprocity (Trivers 1971, Axelrod and
Hamilton 1981, Axelrod 1984). The first answers are biologically
inspired on the increased biological fitness of kinship or on the
adaptive success of groups of Cooperators. Still, the idea of
social kin selection has been put forward (Riolo et al. 2001):
cooperation can arise from similarity, and tags that identify such
similarity can be based in cultural attributes instead of being
genetically determined. Indeed, cultural transmission has been
invoked as a potential reason for the prevalence of cooperation in
human populations (Mark 2002). Other route to cooperation, based
on a reputation score for each individual and information sharing
(Nowak and Sigmund 1998), also emphasizes the cultural forces
present in human society. Cooperation grounded on reputation can
be seen however, as a form of indirect reciprocity.

\begin{figure}
\epsfig{width=0.45\textwidth,file=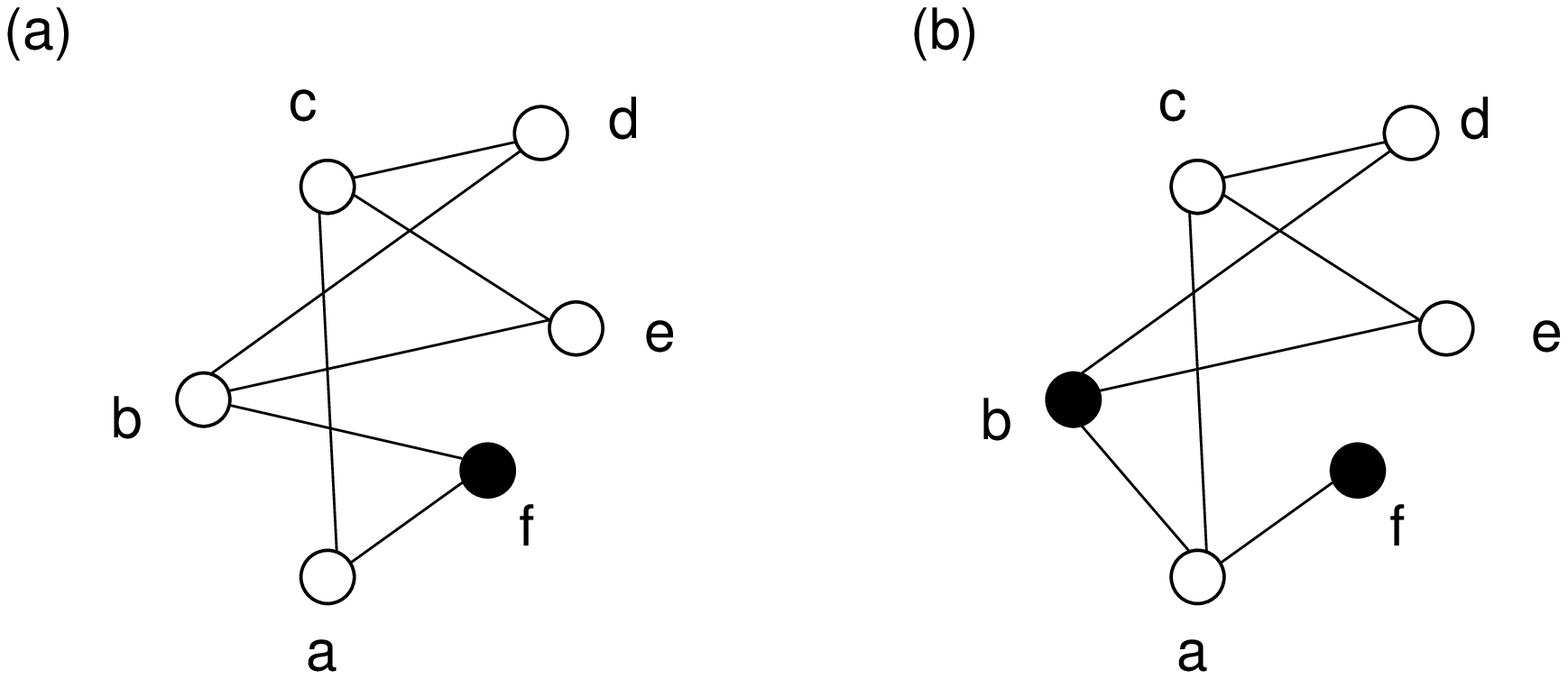} \caption{(a) At time
step $t$, the system is composed by Cooperators (white circles $a$
- $e$) and Defectors (black circle $f$) whose interactions are
given by the links. Given that the payoff matrix is given by
$P=S=0$, $R=1$ and $T=1.3$, the payoff for each agent is
$\{\Pi_a,\Pi_b,\Pi_c,\Pi_d,\Pi_e,\Pi_f\}=\{1, 2, 3, 2, 2, 2.6\}$.
Therefore the Cooperator $c$ is the {\em leader}, because it has
the maximum payoff. Agents $a$ and $b$ get a lower payoff than
their common neighbor, the Defector $f$. However, agent $a$ has
another neighbor $c$ who gets an even larger payoff than $f$. (b)
In the next time step $t+1$, $a$ imitates Cooperation from $c$,
which has the largest payoff among its neighbors, and only agent
$b$ changes strategy and imitates the D-strategy from $f$. The
figure also shows that $b$ broke the link with agent $f$ and
starts a new interaction with the randomly selected agent
$a$.\label{fig0}}
\end{figure}

Cooperation based upon reciprocity is often formalized through the
iterated Prisoner's Dilemma (Rapoport and Chammah 1965). The game
theoretical formulation of the PD shows that two perfectly
rational agents interacting once, and confronted with the choice
of cooperating or defecting, would both choose to defect, which
corresponds to the Nash equilibrium of the game. When the game is
repeated or iterated, the Folk theorem classifies the many
possible outcomes that can be sustained, in particular full
cooperation. It is the {\it shadow of the future} in the repeated
interaction that makes cooperation sustainable (Axelrod 1984).
Although basic concepts and results on the PD were well known in
game theory (Binmore 1998), the evolutionary ideas for the
selection of an equilibrium pioneered by Axelrod (1984, 1997) have
been extremely influential. They have lead to the consideration of
the virtue of different strategies and to the concept of
evolutionary stable strategies. It is now well accepted that
dynamical models of cultural evolution and social learning hold a
greater chance for success than models merely based on rational
choice. This body of knowledge has been reviewed by Hoffmann
(2000), while some recent advances in the theory have been
reviewed by Axelrod (2000).

Generally speaking, the ingredient of locality is not taken into
account in the different studies of models of cooperation
mentioned so far. Individuals or players interact globally by
continuous changes of random pairings among them. However, local
spatial interactions introduce another possibility of reciprocity
which is not based on history dependent or backward-looking
strategies. In this line of thinking, spatial PD games have been
introduced (Axelrod 1984, Nowak and May 1992), and further
generalized by Lindgren and Nordahl (1994). Spatial games and
local interactions were also introduced in economic contexts by
Blume (1993) and Ellison (1993). Nowak and May (1992, 1993) and
Nowak et al. (1994a, 1994b) start considering a group of
individuals placed at the nodes of a regular two dimensional
square lattice. What they showed is that cooperation might arise
from spatially distributed interactions: if agents play only with
a local neighborhood in the lattice, clusters of Cooperators may
survive the interaction with Defectors, and cooperation may be
sustained. This can happen bypassing any consideration on memory
and strategies, in situations in which non-cooperative behavior
would prevail in the global game. Admittedly, the game theory
formulation of cooperation neglects interpersonal relations driven
by emotional processes (Lawler and Yoon 1998). In spatial PD
games, the fundamental relationship is that of exchange
conditioned by structural position. Still, the basic mechanism of
imitation of successful neighbors is introduced. This mechanism is
certainly prevalent in many human interactions. We finally mention
that the analysis of spatial games, specially when searching for
collective emergent behavior in societies with a large number of
individuals, require the use of intensive computer simulations.
There are however some analytical results (Schweitzer 2002),
specially in simplified one-dimensional models (Eshel et al.
1998).

There are other routes to cooperation, among which we can mention
optional participation (Hauert et al. 2002), and stochastic
collusion (Macy 1991). Optional or volunteering participation is a
mechanism that incorporates, in addition of Cooperators and
Defectors, players ({\it loners}) that refuse to participate in
the game. This has been shown to be an effective mechanism to
escape from the social dilemma without invoking any form of
reciprocity. Incorporating locality and spatial interactions,
Cooperators also tend to fare better (Szabo and Hauert 2002).
Stochastic collusion stems from a forward-looking route to
learning and adaptive behavior (Macy and Flache 2002): stochastic
search by adaptive individuals in response to immediate outcomes
allows escaping from the non-cooperative social trap. The
beneficial implications of locality ({\it bounded social space})
to sustain cooperation have also been discussed for this mechanism
(Macy 1991).

\subsection{Social networks}

A new field of opportunities for modeling is opened when
considering the network of social interactions - links between
individuals are not established in a random way, but depending on
neighborhood or interest relationships. As a consequence, clusters
and stable linkage are developed giving way to new aspects of
cooperation in which interactions among individuals have an
important effect, being crucial to the performance of the system
as a whole (Holland et al 1986).

\begin{figure}
\vskip -2.5cm
\epsfig{width=0.5\textwidth,file=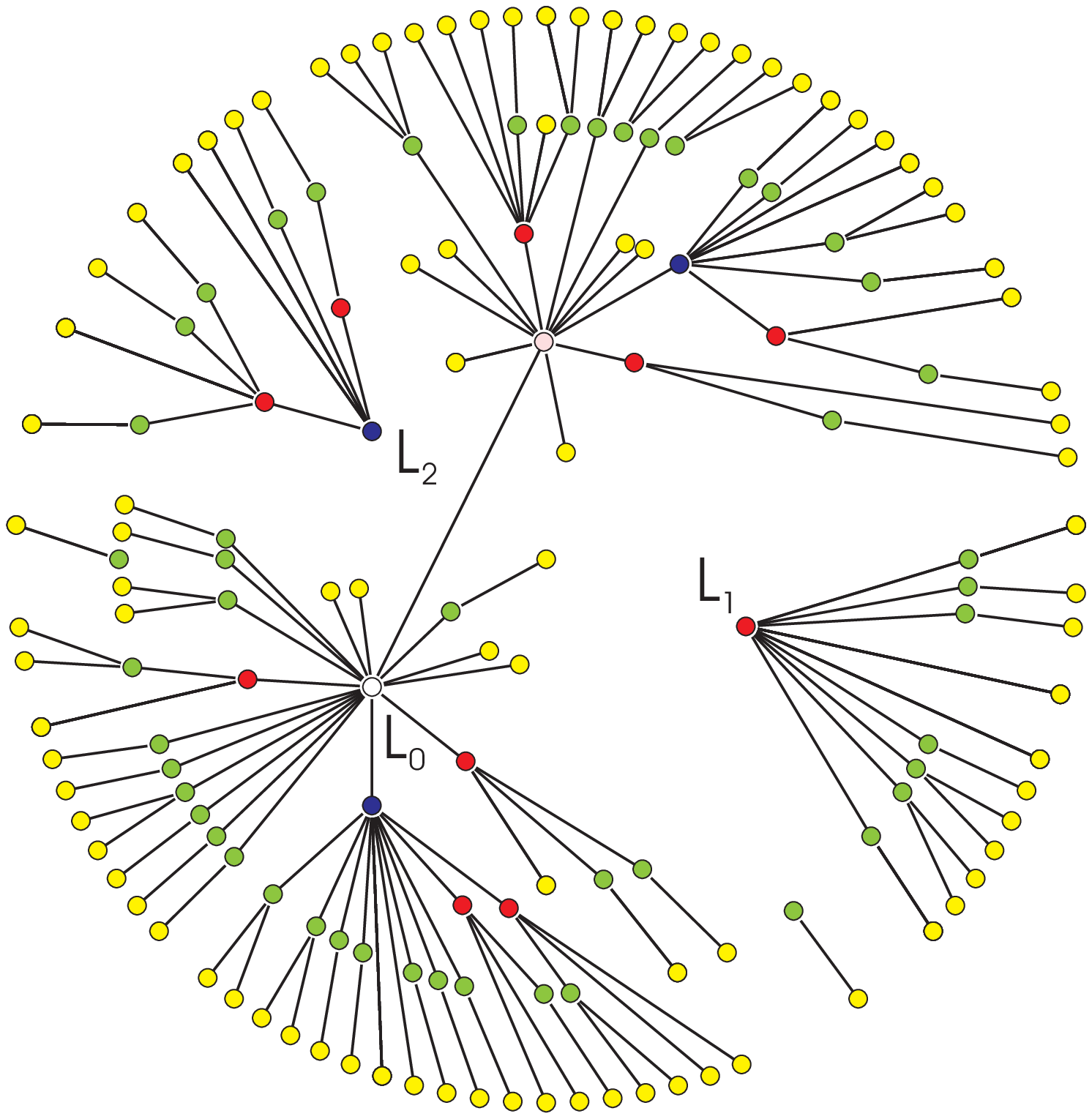}
\vskip -3cm
\caption{Imitation network obtained from a numerical simulation
with $T=1.75$, $R=1$, $P=S=0$ and $p=0.10$. The simulation starts
from a random network with $N=10000$ agents and a number $K=8$ of
average links per node of the network. Agents are organized in
imitation layers coded with the same grey level. Starting from the
outer layer of a tree, new nodes are introduced for each new link
in the path to reach the leader at the top of the tree. Each agent
imitates the agent of the inner layer to whom it is linked in the
tree. The network contains several trees. Three leaders ($L_0$,
$L_1$ and $L_2$) are identified. $L_0$ is the absolute leader with
the largest payoff in the system. Its tree contains approximately
half the number of agents of the system. All nodes are
Cooperators, because Defectors are isolated in the imitation
network. Most of the network's nodes, which belong to the lowest
layer of the hierarchical structure, are not shown for clarity.
\label{fig1}}
\end{figure}

The consideration of the spatial version of the PD discussed above
is a step in the direction of considering a social network. But a
social network is in general different from a regular two
dimensional lattice, as much as social interactions are different
from a continuous random pairing of individuals. The popular
phenomenon of {\it six degrees of separation} (Guare 1990), that
follows from the early experiments of Milgram (1967), makes clear
that if the average separation between two individuals is given by
only six intermediate acquaintances (six links), social networks
radically differ from the regular ones often considered in spatial
games. Such small world effect has been re-popularized by Watts
and Strogatz (1998). Beyond the physical and topological
properties of the network, it is also important to recall that, in
economic terms, a social network has been associated with a social
capital (Coleman 1988), in the sense that it gives the basis for
stable repeated social interactions.

The spatial version of the PD has been revisited by Axelrod and
coworkers (Axelrod et al. 2000, Cohen et al. 2001) trying to
understand the role of social structure and geographically based
networks in the maintenance of cooperation. They have shown that
geographically dispersed social networks are efficient in
maintaining cooperation, provided the links are stable. They
conclude therefore that the important ingredient is not clustering
(i.e., locality or correlation of linkage patterns), but what they
label as context preservation, that is the continuity of
interactions. Another interesting contribution in this context is
the study of Buskens and Weesie (2000) on the role of social
structure in supporting cooperation via reputation. By including a
social structure, this study goes beyond the ones of Nowak and May
(1998) and Riolo et al. (2001) and it shows (Axelrod 2000) that
these two aspects, social structure and reputation, reinforce each
other in building up a cooperative collective behavior. Reputation
established via information sharing leads to sustained cooperation
in social structures that are less rigid than the ones fixed by
geographic positions, adding therefore a higher degree of freedom.

From the early mathematical analysis of random networks (Erdos and
Renyi, 1959) there is a fast growing literature devoted to
uncovering the topological properties of technological, biological
and social networks (Wasserman and Faust 1984, Watts and Strogatz
1998, Watts 1999, Watts 1999b, Barabasi and Albert 1999, Amaral et
al. 2000, Newman 2001, Newman and Strogatz 2001), partially
reviewed by Albert and Barabasi (2002) and Barabasi (2002).
Strogatz (2001) emphasized that, despite their different nature,
many networks present similar characteristics, but it is important
to note that differences do also exist. The main properties of a
network usually analyzed are average distance between nodes,
clustering and degree distribution.  A short average distance,
that is, the fact that the number of steps needed to connect any
pair of individuals in a social network is small, is what
originally characterizes a small world effect. But it was long ago
recognized (Wasserman and Faust 1984) that social networks also
show a high tendency to form cliques, i.e., friends of friends are
also friends. Such high clustering property and a small path
length characterize the small world networks formalized by Watts
(1999). These {\it small world networks}  are half-way between
random and regular networks. The degree distribution is the
probability distribution of the number of links or connections of
a node of the network. Many networks have a power law
distribution, which implies a scale free distribution of degree
(Barabasi and Albert 1999). However, it has been reported (Amaral
et al. 2000) that social networks do not generally display such
heterogeneity in their connectivity, but instead show a more
homogeneous degree distribution, with the number of connections
being characterized by a narrow distribution around a single
scale.

Our concern in this paper is not the characterization of the
topological properties obtained considering a snapshot of a social
network. Rather, the question addressed is how the network is
dynamically formed or how a given network structure is reached
after social agents interact for a long time (Zimmermann et al.
2004). As Watts indicates (Watts 1999), networks affect the
dynamics of the system in a passive and an active way. Examples of
the passive way are the spatial version of the PD game and the
variants thereof previously mentioned, or the study of a PD game
in small world network (Abramson and Kuperman 2001). We are here
interested in the active aspect in which the network of
connections evolves by the will of the agents. We seek to unveil
possible dynamical mechanisms to achieve a small world
connectivity.

\subsection{Making choices and beyond: Spontaneous social
differentiation}

Incorporating the will of the agents seems to be a crucial
ingredient in any model of human behavior. In the setting of a
spatial version of the PD game in a social network, the agents
should have some way of choosing their actions (cooperate or
defect) and choosing their partners. Making choices is an
instrument for learning in an adaptive evolution.

\begin{figure}
\epsfig{width=0.4\textwidth,file=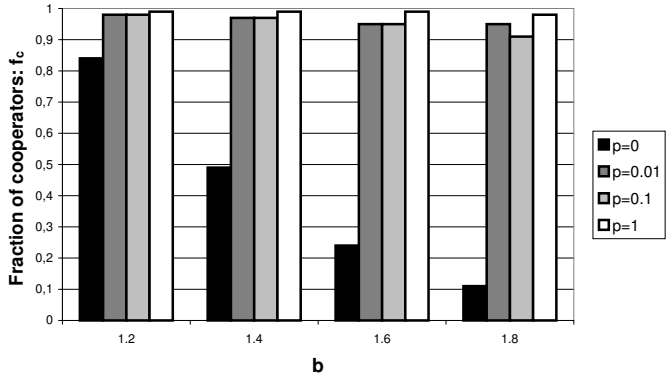,angle=90}
\vskip -1.5cm
\caption{Average fraction of cooperative agents, $f_C$, for
different values of the plasticity $p$ and incentive to defect,
$b$. For a fixed network ($p=0$), $f_C$ decreases with $b$.
However in the presence of social plasticity ($p \ne 0$) $f_C$ is
kept above 90\%. \label{fig2}}
\end{figure}

Partner and action selection has been taken into account in the
iterated PD as reviewed by de Vos et al. (2001), but without
considering local spatial interactions and also without reference
to a social network. The volunteering mechanism of Hauert et al.
(2002) does not give a choice of action, since agents refusing to
participate are fixed from the outset, but it is an indirect
mechanism to determine who interacts with whom. More closely
related to changes in the network structure, or to the choice of
partners, is allowing players to exit from an unsatisfactory
relationship with partners, as discussed by Axelrod (2000) on the
grounds of the Edk-Group's analysis of a set of fifteen strategies
related with the possibility of opting out (Edk-Group 2000). The
possibility of exiting generally reinforces cooperation, but again
this study does not include local interactions and social
networks.

The formation of a social network based on individual decisions
has been studied by Bala and Goyal (2000), but differently of our
interest here, the individuals forming the network do not have
some dynamics on top of the network of interactions: there is no
game mediated by the interactions defined by the network. Skyrms
and Pemantle (2000) do consider the simultaneous evolution of
actions and structure of the network (fluid network), emphasizing
that these type of models are largely unexplored. A generic result
from their study is that important changes of global behavior
occur when moving from frozen to fluid social networks. Their
approach is similar to the one also considered by Macy (1991):
Starting from a random network, the network is regenerated at each
time step of the game by random pairings among the agents, but the
probabilities of a pairing depend on the previous results of the
game. This a basic idea of co-evolution, but it does not take into
account aspects of locality, or a social network, that albeit
evolving, has well established stable links among individuals.

Our starting point is the observation that the influence of social
structure in human behavior, with feedback between actions and
social structure, might be seen as an entangled co-evolution of
the choice of actions and the formation of a social structure.
Specifically, we address here the issue of the evolution of social
networks in the context of cooperation, allowing individuals both
to choose actions and to choose exiting from unsatisfactory
relationships. We propose a simple model where the actions of
individuals that form the social network are driven by their level
of satisfaction, choosing their actions by imitation of their best
neighbor. Moreover, we introduce {\em social plasticity} --other
definitions are introduced by \citet{Lazer01}-- as the capacity of
the individuals to choose partners, being able to change their
neighborhood as time goes on. Evolving interactions, thus, appear,
allowing individuals to choose different partners on the grounds
of the obtained performance.

The significant step forward in the results of our approach is
that it naturally leads to a process of social differentiation.
Starting from random partnership among equivalent individuals, a
social structure emerges. In this emerging structure the topology
of the network of social relations identifies individuals with
different social roles. These roles have been spontaneously
selected during the complex adaptive dynamics of the social group.
They are not the direct consequence of initial differences in
strategy or location. Rather they emerge in a probabilistic
dynamics in which the social network is constructed.

\section{Prisoner's Dilemma in an adaptive network}

The simplest form of the PD game consists of two agents which may
choose from either of two actions: to cooperate (C)
or to defect (D). If both agents choose C, each agent gets a
payoff (reward) $R$; if one defects while the other cooperates,
the former gets a payoff $T$ (with $T>R$), while the latter gets
the "suckers" payoff $S$ (with $S<R$); if both defect, both get a
payoff $P$. Under the standard restrictions $T>R>P>S$, $T+P<2R$,
defection is the best choice in a single shot game (Nash
equilibrium). Thus the social dilemma of how cooperation may be
sustained arises, due to the fact that rational agents would
defect. When no social structure is assumed, agents are drawn
randomly in pairs to play the game, and the dynamics may be
described by a replicator type equation \citep{Hofbauer98}. This
equation can be understood from the biologically motivated fact
that strategies with fitness larger than the average in the whole
population, replicate with a positive rate, while those that
underperform the average, die away. In our context, if $\Pi_C$ is
the average payoff obtained using strategy C, and $\langle \Pi
\rangle $ is the average payoff for all the population, then the
time evolution of the fraction of the agents using strategy C,
labeled $f_C$, obeys the equation
\begin{equation}
\frac{df_C}{dt} =  f_C (\Pi_C - \langle \Pi \rangle)~.
\end{equation}
Given the payoffs of the PD game we obtain
\begin{eqnarray}
\Pi_C &=& R f_C + (1-f_C) S \\
\langle \Pi \rangle &=& R f_C^2 + (S+T) f_C (1-f_C) + P (1-f_C)^2~,
\end{eqnarray}
and thus
\begin{equation}
\frac{df_C}{dt} =  f_C (1-f_C)((R-T)f_C + (S-P)(1-f_C))~.
\end{equation}
The analysis of this equation indicates that it has two
equilibrium solutions in $f_C \in [0,1]$: $f_C=0$ is a stable
fixed point, while $f_C=1$ is an unstable fixed point. Thus, it
also follows from this dynamical analysis that in the absence of
an interaction network describing a social structure the only
stable solution is a pure defective state.

\begin{figure}
\epsfig{width=0.45\textwidth,file=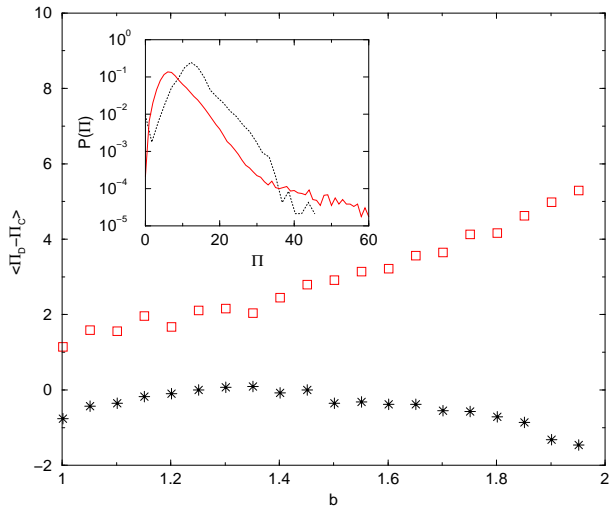}
\caption{Difference between the average payoff of Defectors,
$\langle \Pi_D \rangle$, and Cooperators, $\langle \Pi_C \rangle$,
as a function of $b$ for $p=0$ (stars) and $p=0.10$ (squares). The
effect of the social plasticity is that the Defectors get a larger
payoff on average: $\langle \Pi_D -\Pi_C\rangle$ is positive for
$p\ne0$. Inset: probability distribution (normalized to each
sub-population) of individual payoff, for Cooperators (solid line)
and Defectors (dotted line). The most probable payoff, which
corresponds to the maximum of the distribution, is larger for
Defectors than for Cooperators. However, there are a few
Cooperators getting the largest payoff in the population as it is
seen in the tails of the distribution for large values of $\Pi$.
Parameter values $b=1.75$, $p=0.1$. \label{fig3}}
\end{figure}

In the spatial version of the PD
\citep{Axelrod84,Nowak92,Nowak93,Nowak94a,Nowak94b} agents sit on
the nodes of the network and they play the game with their
neighbors. Two agents are said to be `neighbors' if they are
directly connected by a link of the network. A detailed analysis
of the different spatiotemporal patterns that can arise in the
dynamics shows a phase space where Cooperators and Defectors can
coexist depending on the parameter values
(\citep{Schweitzer02,Lindgren94}). There are threshold values in
parameter space beyond which defection dominates the whole
network.

We will consider a spatial version of the PD, generalized in order
to consider, instead of fixed, regular lattices, {\it adapting
evolving networks} of interaction.

\subsection{Social plasticity}

We want to study how the results obtained with fixed interaction
networks change if certain {\em social plasticity} is allowed.
This term addresses the common observation that social
interactions in most societies adapt in time by a learning
process. We will implement  a rule for social plasticity which
depends on both the strategies and pay-off of the individual
players. The motivation of this rule comes from the following
analysis of the mutual benefit obtained from each pair of allowed
PD interactions: two symmetric (C--C and D--D) and one asymmetric
(C--D). Clearly the symmetric interaction C--C should be
reinforced because both agents get the maximum payoff from their
selected strategies. However, the opposite occurs in a D--D
interaction, where both agents have aligned incentives to change
neighbor and to possibly find a C-neighbor. The asymmetric
interaction C--D forms an intermediate class, where the C agent
will not support the interaction, but the D agent will try to
reinforce it. As a first approximation to the problem, we assume
that this type of interaction does not change, because the overall
effect of both agents is balanced. From this analysis, we conclude
that the simplest non-trivial social plasticity rule should allow
interactions among two D-agents to adapt, providing a
self-interest mechanism of social adaptation where the agents can
increase their payoffs by changing their partners.

\subsection{The model}

We consider a fixed population of agents placed in the nodes of a
network and connected by links to their neighbors. The dynamics
evolves in discrete time steps divided in three stages, starting
from a random choice of strategies and a random network:
\begin{enumerate}
\item {\em Interacting}: Each agent $i$ plays the PD game with its
neighbors, and collects an aggregate payoff $\Pi_i$;
\item {\em Strategy update}: Each agent updates its current
strategy, imitating the strategy of the neighbor with the largest
payoff including himself (whenever more than one equivalent
neighbor with a larger payoff exists, one of them is randomly
selected);
\item {\em Neighborhood update}:
If agent $i$ imitates a Defector, then the agent $i$ replaces with
probability $p$ this link with the imitated D-neighbor by a new one
pointing to a randomly chosen partner from the whole network. This
process updates the network.
\end{enumerate}

In order to clarify the rules of the model we
would like to remark the following points.
First, we specify that in the interaction step we only consider
bidirectional (undirected) links, so if agent $i$ plays with $j$ then we also
assume that $j$ plays with $i$.
Second, we do not consider complex
strategies which involve the history of past encounters with
neighbors (as in \citet{Lindgren94}), but we consider instead only
zero-memory strategies, and assume that each agent plays the same
strategy (action) C or D with all its partners.
Third, the game is played synchronously, i.e., at each discrete
time step agents decide their strategy in advance and they all
play at the same time.

\begin{figure}
\epsfig{width=0.5\textwidth,file=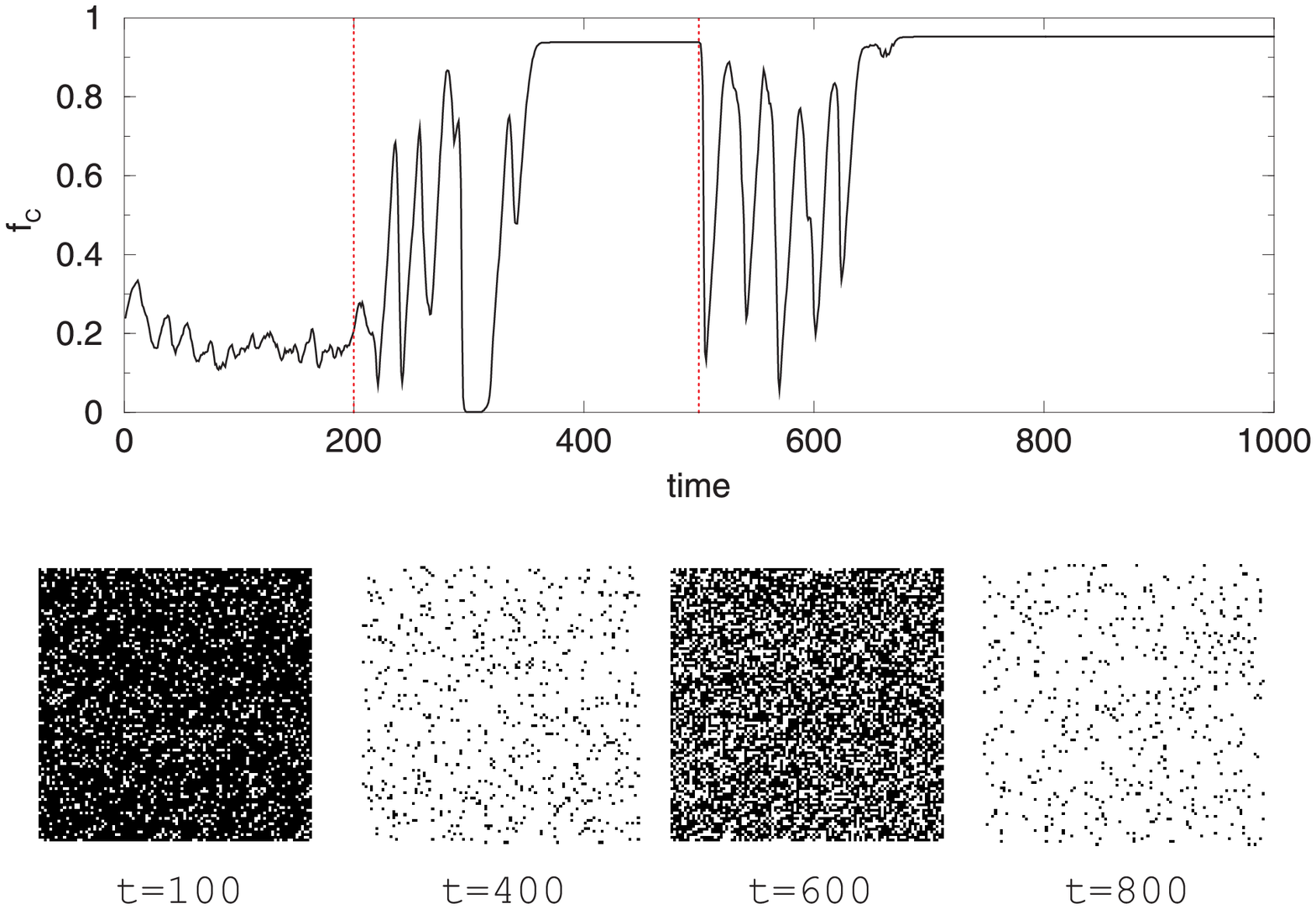}
\vskip -6cm
\caption{Time series for the fraction of Cooperators $f_C$ for
$b=1.7$. The evolution is in a fixed random network ($p=0$) up to
time $t= 200$ when network dynamics is switched-on, so that $p=1$
for $t>200$. At time $t=500$ the network leader is forced to
change action from C to D, to show how cooperation may recover
after a social crisis. After an oscillatory transient, a new
stationary state is reached. Below, snapshots of the instantaneous
configurations of actions are shown at the indicated times. A
white (black) point is a C (D) agent in a node of the network
(links are not shown). The spatial location of the agents does not
have any meaning due to the building of a network of interactions.
It is only used for displaying purposes. \label{fig4}}
\end{figure}

We consider in Fig.~\ref{fig0} an example that illustrates the
rules of the model. The system is initially formed by 5
Cooperators (agents $a$ to $e$) and 1 Defector (agent $f$).
According to the payoff they obtain, agents $a$, $d$ and $e$ are
unsatisfied and they will imitate in the next time step the
Cooperator $c$. Agents $c$ and $f$ are satisfied because they
don't have a neighbor with a larger payoff. Finally agent $b$ is
also unsatisfied but the agent with largest payoff is the Defector
$f$. Thus in the next time step it changes its strategy to become
a Defector, and (with probability $p$) breaks the link with agent
$f$ and selects a new neighbor as partner (in this case he selects
agent $a$). Note that in the neighborhood update rule it is not
needed that an agent changes its strategy from Cooperator to
Defector to severe a link. This example shows that the
neighborhood update rule facilitates the survival of the Defectors
by increasing their payoff when a Cooperator is selected at
random.

The strategy and neighborhood update rules implemented in this
model represent a useful choice to unveil basic mechanisms behind
global cooperative behavior. Other choices are discussed later in
Section~4. On the one hand, the extreme learning (imitation) rule
of the strategy update can be justified by the psychological bias
to focus on confirmation and neglect disconfirmation of believes
(Strang and Macy, 2001). Bounded rational agents seek to learn
from limited and biased information: they respond to perceived
failure by imitating their most successful peer. In our case
example of scientific collaborations, each scientist takes as a
model to imitate the most successful among his collaborators
rather than, for example, alternating in imitating some of them.

Regarding the neighborhood update rule, we assume first that only
D--D links are broken, and second that any targeted agent always
accepts a new partner. This rule can be justified  as a
conservative assumption when considering minimal conditions for
the emergence of cooperation. As discussed in Section~4,
alternative rules allowing Cooperators to severe their links would
further enhance cooperation. The second assumption can be
justified invoking the absence of cost to sustain a link. Within
this assumption, the new link may only increase the payoff, never
decrease it, thus it will always be accepted. In the context of
scientific collaboration, if the cost to sustain a research
collaboration was zero, then scientists would accept offers from
any collaborator with the expectation to get a benefit. Beyond
sustaining a link, there is a cost to establish a new on. In our
model, the plasticity parameter $p$ measures how easy is both to
severe a collaboration and find a new partner. While for $p=0$
this cost is extremely large (so large that it prevents the
formation of new links), for $p=1$ the cost is small.

Finally, although in real social dynamics one experiences both
spontaneous creation and suppression of interactions, for
simplicity, our model assumes that unsuccessful relations are
replaced by randomly selected new ones, leaving always the total
number of links constant.

Our strategy update introduces a first classification of the
agents in the network: {\em satisfied} and {\em unsatisfied}
agents. An agent is satisfied if he does not imitate any other
agent, so he has the largest payoff in its neighborhood; otherwise
he is unsatisfied. Thus, strategy {\em imitation} and {\em social
plasticity} (adaptation) is restricted only to unsatisfied agents.

The probability $p$ measures the social plasticity of the agents,
controlling the rate at which the network structure evolves, as
compared to the time scale of evolution of the strategies. For
values of $p \ll 1$ strategies change much faster than network
evolution (situation close to the frozen network of $p=0$), while
for $p=1$ strategies and network evolve at the same rate (fluid
social network).

\subsection{Stationary states}

The proposed model naturally leads to a time evolution of the
local connectivity of the network, featuring agents with
heterogeneous neighborhoods. Starting from initial condition of
random partnership with links randomly placed among initially
equivalent agents, the feedback between choice of strategies and
choice of partners results in a dynamical evolution from which a
well-defined social structure is expected to emerge in the form of
{\em stationary states}.

\begin{figure}
\epsfig{width=0.4\textwidth,file=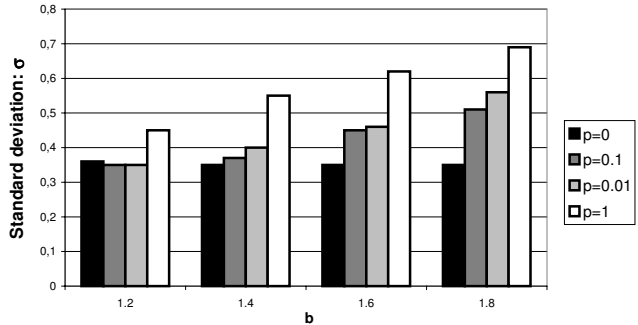,angle=90}
\vskip -2cm
\caption{Normalized standard deviation $\sigma_n$ of the degree
distribution of the social networks obtained for different values
of the plasticity parameter $p$ and the incentive to defect $b$.
For a Poisson distribution and $K=8$, $\sigma_n = K^{-0.5}=0.35$,
while for an exponential distribution $\sigma_n = 1$. For $p=0$,
$\sigma_n$ takes the value of a Poisson distribution. As $p$ and
$b$ are increased the interaction network departs from a Poisson
distribution and approaches a distribution with an exponential
tail. \label{fig5}}
\end{figure}

In our context a stationary state is reached when both the
strategy of each agent and its neighbors remain fixed in time.
Given the discrete nature of the dynamics and the finite number of
possible states, these states may be reached in a finite number of
time steps. The simplest stationary state consists of all agents
being Cooperators. Note however, that a state composed exclusively
by Defectors is a stationary state only when the network is fixed
($p=0$); for $p>0$, the agents strategies remain fixed, while the
network is continuously evolving. According to the dynamical rules
of the model, each agent being a Defector is not a stationary
configuration, even in the case of a social network where each
agent has the same number of neighbors (remember than in the case
of several neighbors sharing the largest payoff in the
neighborhood one of them is picked at random). In an all-D network
only interaction links change, but not strategies. For the sake of
clarity we will refer to this state as the all-D network state,
even if the interactions are not stationary. Our system has a
multiplicity of different stationary states and it is expected
that the system reaches one of these states. This assumption is
confirmed by numerical simulations. The specific stationary state
reached by the system depends on the stochastic process that
shapes the network evolution. A proper characterization of the
general properties of these stationary states relies then on
statistical tools.

Two requirements are needed for a stationary state to exist. The
first is that there are no links between two Defectors, so that
the neighborhood update rule ($p>0$) does not produce any change.
The second condition relates the respective payoffs in each
neighborhood of any Cooperator $i$ interacting with a D-agent, say
$d$.  Their respective payoffs must satisfy
\begin{equation}
\Pi_j > \Pi_d > \Pi_i~,
\label{Pi}
\end{equation}
where agent $j$ is the Cooperator imitated by $i$. Thus, a
stable situation occurs when the payoff of Defectors accommodate
'in-between' the payoff of two Cooperators, and they only {\em
exploit} Cooperators. Whenever the payoff of $d$ becomes larger
than the best neighbor of $i$, the configuration is no longer
stationary, because the Cooperator $i$ will switch to imitate the
strategy of agent $d$. Thus, in the stationary state, Defectors do
not interact with other D-agents, and no Cooperator imitates their
strategy. An interesting consequence of this result is that the
agent with the largest payoff is a Cooperator, if the system
settles to a stationary state. However, this does not prevent that
in a transient state a Defector can hold the largest payoff.

This simple analysis highlights the role that the `imitation of
the best neighbor', coupled to the social plasticity rule, has in
shaping the social structure of the system. A related consequence
is that the social structure developed becomes hierarchical in
terms of payoff and in terms of the dynamics as we will show
below. One way to visualize the hierarchal structure of the
network is by constructing a sub-network referred as {\em
imitation network}, where each {\em directed} link indicates who
is imitating whom. As in stationary states no other agent imitates
Defectors, in this representation, D-agents are isolated.

Figure~\ref{fig1} shows part of the imitation network in a
stationary configuration. The agents in the outer layers imitate
the ones in the inner layers: by our previous definition they are
unsatisfied agents. In a stationary state, Cooperators form {\em
trees} of agents hierarchically ordered according to their payoff.
This description highlights some special agents that do not
imitate any other agent: the agent in the top of the chain has the
highest payoff of the tree it belongs and he is the only agent
that is satisfied in the tree. They are easily recognized at the
top of the trees. All other agents in the tree are unsatisfied,
but they imitate the same strategy they were playing in the
previous time step. An implication of this representation is that
the size of the {\em group} formed by following directed links to the
leader, gives an indication of the influence of the leader with
respect to the rest of the system. In a stationary state, a leader
has no links to D agents. Among the leaders there is an absolute
leader defined as the one with the largest payoff in the system.
As a consequence it is also the agent with the largest number of
links in the imitation network. All the links of the absolute
leader are imitation links shown in the imitation network; any
other leader in the system has a smaller number of cooperating
partners. The whole system is dominated by the few leaders that
control a large fraction of the population. The structure of the
social network is then expected to be very sensitive to
perturbations acting on those leaders.

When the system is close to a stationary state (i.e., if there are
a few Cooperators not satisfying Eq.~(\ref{Pi}), as it happens
when, for example, a Cooperator imitates a Defector), then the
imitation network is useful to understand how this 'perturbation'
will propagate along the rest of the social structure. Note that
at each time step, the D strategy replicates on all those agents
connected to the agent where the perturbation started, causing an
'avalanche' of replication events. This chain of events may end in
a all-D network or due to the existence of Cooperator leaders,
cooperation may recover, as we will show below.

\subsection{Social differentiation}

Our previous description of the resulting stationary states,
offers also a description of how our social model with network
adaptation naturally gives rise to a process of social
differentiation, with the spontaneous emergence of different
social roles. Even if all the agents are driven by the same
dynamical rules and they are initially statistically equivalent,
their role in the network diversifies. Our previous analysis of
the imitation network allows us to identify three types of agents
in a stationary state:
\begin{itemize}
\item {\em Leaders}: satisfied Cooperators that have the maximum
payoff in their corresponding neighborhood. The absolute leader is the
agent with the largest payoff in the whole network, and its
corresponding group of influenced agents is in general the
largest. Typically the other leaders have also a large number of
links and a payoff above the average. Admittedly, the sociological
concept of leader has several different characteristics, not all
of them included in our definition. Still we use this term to
emphasize that those are agents strongly influencing other agents
to adopt their strategy. In this context leaders do not have the
will of seeking their leadership nor do they try to preserve it.

\item {\em Conformists}: unsatisfied Cooperators, i.e. they do not
have the maximum payoff in their
neighborhood, but they imitate another agent playing their same
strategy. They constitute the large majority of the nodes of the
imitation network except for the leaders.

\item {\em Exploiters}: Defectors who take advantage of other's actions.
Defectors have a larger payoff than their (C) neighbors, showing
they succeed in exploitation, and are satisfied. In the imitation
network they do not have any directed link, because in the
stationary state no other agent imitates their strategy.
\end{itemize}

\section{Simulating social dynamics}

To facilitate comparison with previous results of the PD game in
fixed networks (e.g.., \citep{Nowak92}), we have performed
numerical simulations of the above model using as main parameters
the incentive to defect $b\equiv T$ in the range $1<b<2$ and the
social plasticity $p\in[0,1]$. The rest of the PD payoff matrix
parameters have been kept fixed, $R=1$, $P=0$, and $S=0$ as in
\citet{Nowak92}. It has been previously found \citep{Lindgren94}
that turning $P=S=0$ does not change the main results, although
the single PD game is not in the strict Prisoner's Dilemma
conditions. We have checked that changing the parameter $P$ in the
region $(0,0.1)$ does not affect significantly our results.

\begin{figure}
\epsfig{width=0.5\textwidth,file=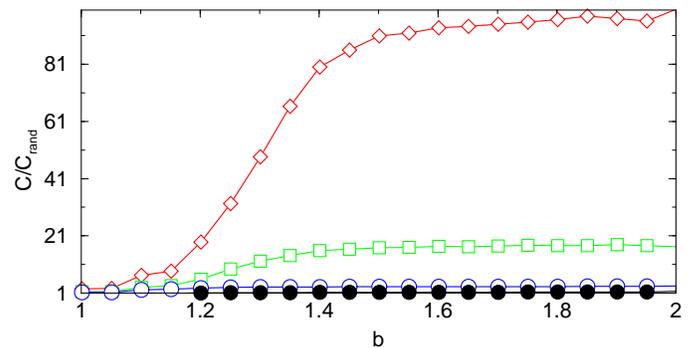}
\caption{Normalized clustering coefficient, $c/c_{rand}$, with
$c_{rand} = K/N$. Filled circles correspond to $p=1$ and $q=0$.
Other curves are for $q=0.01$ and $p=0.01$ (diamonds), $p=0.1$
(squares), and $p=1$ (empty circles). Social plasticity ($p\ne0$)
and a local selection of new partners ($q \ne 0$) is needed in
order to reproduce the clustering observed in social networks.
\label{fig6}}
\end{figure}

All simulations were performed with $N=10000$ agents. The strategy
initial condition was always set to 60\% of randomly distributed
Cooperators. The network initial condition was set by distributing
$KN/2$ undirected links between random pairs of nodes. The number
$K$ corresponds to the average connectivity per agent, and we
considered $K=8$. The initial degree (number of links of each
agent) distribution constructed in this manner is a Poisson
distribution with parameter $K$. Simulations with other values
parameters have been performed and no qualitative differences were
found. Likewise we have checked that using an asynchronous
updating in our simulations does not change any meaningful
qualitative result. For the above parameters, a stationary state
is reached in a time scale that depends very much on $p$,
requiring about a $1000$ time steps for $p=0.01$ and about $10$
time steps for $p=1$. For smaller $N$ the typical time to reach a
stationary state decreases, while for larger $K$ it takes longer
time. A proper characterization of the general properties of these
stationary states relies then on averaging over many different
numerical experiments. Our results are given as averages over
$100$ different experiments.

Simulations in our adaptive dynamic network game show how
Cooperation is in general enhanced and no threshold to an all-D
network is observed. In fact, our results indicate that for a
given set of parameters, there is a {\em coexistence} of a
multitude of cooperative stationary states and the all-D network
state. This means that for a fraction of the initial conditions,
the stochastic dynamics of the network may lead either to the
all-D network, or to a stationary state. Once in the all-D
network, the system gets trapped and no recovery is possible
(remember that an in all-D state the interaction links change, but
not strategies). The probability of reaching the trapping all
D-network, decreases with increasing the number of agents (it also
decreases with decreasing the incentive to defect $b$ and
increasing the initial distribution of Cooperators), and thus the
trap appears to be a finite size effect. For instance, for an
incentive to defect $b=1.6$ and plasticity $p=0.01$, only 6\% of
the realizations get trapped for a large population of $N=10000$,
while in contrast for $N=1000$, 80\% get trapped. Here we
concentrate on stationary states where Cooperators coexist with
Defectors.

\subsection{Fraction of Cooperators}

Figure~\ref{fig2} shows a first global characterization of the
stationary states, for different values of the parameter $p$ and
the incentive to defect $b$. The fraction of Cooperators, $f_C$,
is measured for a fixed network ($p=0$) and is shown to decreases
with $b$, approaching 0 at a threshold value of $b\simeq 1.75$.
Thus context preservation \citep{Cohen01} without social
plasticity provides partial cooperation. In clear contrast, for
positive $p$, social plasticity facilitates the establishment of a
highly cooperative state with a fraction of Cooperators
essentially independent of the incentive to defect $b$. Even for
small plasticity $p$, the fraction of Cooperators is above 90\%
in the range of parameter $b$ considered. A similar result is
obtained for other values of the average connectivity $K>2$. In
this sense, context preservation is not a necessary condition to
build up cooperation, but rather the social structure is a
consequence of an adaptive dynamics in which cooperation is
greatly enhanced.

It is worth mentioning that the result of enhanced cooperation
works against the direct effects of the neighborhood adaptation
rule, introduced to facilitate the survival of Defectors.
Defectors are allowed to find new Cooperators to exploit, exiting
from unsatisfactory interactions with other Defectors. However,
the final outcome is that the number of Defectors decreases. It
seems that successful Defectors become isolated because they are
role models: their victims ``run away''. The new Defectors --those
that imitated a successful Defectors--  establish links with some
Cooperators who have a high concentration of some other
Cooperators in their neighborhoods. Thus the formerly exploited
Cooperators, now Defectors, turn again into Cooperators. What it
is needed to get cooperation is Cooperators with enough Cooperator
partners to have a payoff higher than any Defector. This
hypothesis is analyzed in the following sections.

\subsection{Distribution of Cooperators' and Defectors' payoffs}

The total average payoff is larger in the dynamic network
($p\ne0$) than for a frozen network ($p=0$). However, we have
measured systematic differences when considering the average
payoffs $\Pi_D$ and $\Pi_C$, for each of the respective
sub-populations D and C. While the number of Cooperators is
larger, their average payoff is smaller than the one of the
Defectors, reversing the situation of what happens in a frozen
network (see Fig.~\ref{fig3}). The payoff distribution in a
stationary state for each subpopulation is shown in the inset of
Fig.~\ref{fig3} for a particular value of $b$. This graph shows
that the most probable payoff is larger for Defectors, explaining
also the larger average payoff of the Defectors. However, close
inspection reveals that the distribution for Cooperators has
always a larger tail, indicating that there is a number of
Cooperators with a payoff larger than any other Defector in the
network. Most of these agents belong to the class of the leaders,
and are necessary for a cooperative stationary state to exist.

Thus the main characteristics of our adaptive social game, is that
an altruistic social network, i.e., a network composed by a large
fraction of Cooperators, may develop with few Defectors but
which have a larger average payoff. The neighborhoods of the
agents adapt to conform the requirements of the stationary states
(D-agents only interact with Cooperators), and the Cooperator with
a largest payoff in each branch of the imitation network
corresponds to a `leader', with a payoff much larger than the
average, from which all other agents imitate its strategy.

\subsection{Transient dynamics, leaders and social crisis}

The evolution towards a stationary state is typically not a smooth
monotonous build-up of a globally cooperating state. Starting from
a random initial condition, the transient dynamics is
characterized by small fluctuations and some distinctive large
oscillations in the fraction of Cooperators as a function of time,
that can be understood in terms of {\em social crisis}.

An example of the frustrated attempts to built cooperation is
given by the evolution of the fraction of Cooperators shown in
Fig.~\ref{fig4}. Starting from a non-stationary state of low
cooperation in the initial fixed random network for $t<200$,
network dynamics then leads to a social network with a high degree
of cooperation after several large oscillations in the time
interval $220<t<300$. The large oscillations in $f_C$ are
frustrated attempts to build cooperation. This indicates that the
defecting behavior is so rewarding, that the cooperation has to
find a specific network configuration in order to be robust
against eventual changes of strategy. In such configuration the
most connected agent (largest number of links) in the imitation
network is also the one with largest payoff. In the frustrated
attempts to reach a global stationary cooperative state, the
fraction of Cooperators becomes large. A few time steps before
$f_C$ reaches each maxima, a Cooperator receives a new link from a
Defector with larger payoff, and switches to D-strategy, starting
a social crisis. Given that at those time steps there is a large
proportion of Cooperators in the network, a fast avalanche of
imitation of D-strategy starts. This imitation will certainly
affect the neighbors of a Cooperator leader, thus decreasing the
payoff of most leaders. Precisely when an avalanche starts, there
is a Defector with a payoff {\em larger} than the {\em absolute
Cooperative leader}. The social crisis propagates through most of
the system, producing a change in the connectivity of the social
network (\citet{Zimmermann01}). The initial distribution of the
number of links for the C and D populations are Poisson
distributions around the average connectivity, typical of random
networks. At the time of a local maximum of $f_C$ the two
distributions have exponential tails for large values of the
number of links. Then very rapidly after a local maximum of $f_C$,
the network switches to the almost defective solution with a large
number of D--D links and Poissonian degree distributions. However
the existence of a small number of Cooperators with a large payoff
permits, thanks to the plasticity of the network, the gradual
build-up of cooperation by creating C--C links. This requires two
steps. First D--C links are created (initiated by the Defectors)
and then the Defectors imitate  the more successful Cooperator in
a later stage. Finally in the stationary state, the distribution
of the number of links for the D population becomes very narrow,
while the distribution for the C population displays a tail
approaching an exponential decay. The stationary network
configuration is thus dominated by a few Cooperators --the
leaders-- with a large number of links (the tails of the
distribution of the number of links for Cooperators). These highly
connected agents dominate the collective behavior of the network.

This characteristic dynamics reveals the {\em functional property}
of C-leaders; on one hand, due to the imitation process they
sustain cooperation, while on the other they enhance cooperation
whenever the C-leader has the largest payoff of the whole network.
In fact there is a sort of competition between the Cooperative
leader and the Defector with largest payoff. We remark that the
latter process is due to the social plasticity: whenever a
Defector selects a leader for partnership, there is a large
probability that it has a larger payoff and the Defector will
imitate cooperation, {\em enhancing} the number of Cooperators.

Figure~\ref{fig4} also illustrates the sensitivity of the
stationary network structure to exogenous perturbations acting on
the leaders, which reflects their key role in sustaining
cooperation. At time $t=500$, the system already reached a
stationary state. However, at this time step, we have forced a
strategy switch of the cooperative absolute leader, leading once
again to a social crisis similar than the previously described in
the endogenous dynamics. In summary, the system self-organizes
itself in one of several possible cooperative states where
avalanches and social crisis are likely to occur following
spontaneous focused local perturbations.

\subsection{Structure of the social network}

A first characterization of the topology of the stationary network
reached by the dynamics is given by the degree distribution, i.e.,
the probability distribution of the number $K$ of links of a node.
This distribution has a long tail that distinguishes it from the
Poisson distribution of the random network. This is reflected in
the values of the normalized standard deviation $\sigma_n = \sigma
/ K$ shown in Fig.~\ref{fig5} for different values of $p$ and $b$.
We recall that for a Poisson distribution $\sigma_n=K^{-1/2}$,
while for an exponential distribution $\sigma_n=1$. We find that
for small values of $b$ the distribution departs significantly
from the Poisson distribution only for large values of the
plasticity parameter $p$, while for increasing $b$ the tail of the
distribution expands and approaches an exponential form. In other
words, the hierarchical structure of the network accentuates as
$b$ increases, with fewer leaders that have a larger payoff.

The distribution of payoffs (inset of Fig.~\ref{fig3}) is similar
to the degree distribution, because the payoff of the Cooperators
is given by the number of links with other Cooperators, and there
is a very small number of Defectors in the stationary state.
Therefore, the variations of $\sigma_n$ also provide a measure of
social inequality in the network. In fact $\sigma_n$ is closely
related to the Gini coefficient \citep{Kakwani80} used in the
characterization of economic inequalities in a social group. We
have measured the Gini coefficient of the resulting network
configuration in a stationary state, and found that it can be
twice as large compared to a random (Poissonian distribution)
network. Social plasticity generates a flux of payoff towards
richer individuals.

Finally, we address the question of whether the social structure
generated in our dynamical model has the characteristics of Small
World network. Two requirements have to be fulfilled: the
clustering (or cliquishness) has to be much larger than in a
random network, and the average path length between two nodes
should be similar to the one of a corresponding random network.
The clustering coefficient, $c$, measures the fraction of
neighbors of a node that are connected among them, averaged over
all the nodes in the network. Most real complex networks show a
clustering larger than in random networks given by $c_{rand}=K/N$
\citep{Amaral00}. In our original formulation, numerical
simulations show (Fig.~\ref{fig6}) that for increasing $b$ the
clustering coefficient increases very mildly with respect to the
clustering of a fixed random network (i.e. $c/c_{rand}$ may change
up to $1.06$). We have tested a slight enhancement of the network
adaptation, which easily accounts for high clustering. Very often,
new acquaintances are made based on the relationships of current
neighbors. To implement this idea, the social neighborhood
adaptation step of our original formulation was augmented: if a
neighbor is replaced by a new one, with probability $q$ a {\em
local} selection of a new partner is done among the neighbors of
the neighbors, while with probability $1-q$ the previous random
selection is performed. The limiting case $q=0$, corresponds to
our original formulation, while $q=1$ corresponds to the case that
all the new partners are chosen from the neighbor's neighbors. It
is natural that this new mechanism will increase the clustering.
Numerical simulations show that while most of our results
previously discussed are qualitatively independent of the value of
$q$, with a very small value of $q$ the clustering coefficient
reaches a very large value. For instance, 1\% of local partner
selection is enough to increase $c$ a hundred times, being the
clustering largest for a slow evolution of the network ($p \ll
1$). For the second requirement for small world topology, we find
that the average path length remains in all cases very close to
the one of a random network. All together our results indicate
that allowing for local partner selection, the social network
generated in our adaptive dynamics has the structure of a small
world network.

\section{Discussion}

We have presented a minimal model which incorporates simultaneous
and coupled evolution ({\it co-evolution}) of the strategies of
the agents and of the social network, providing a first step in
the investigation of the processes of social differentiation in a
globally cooperating social group. Different agents end up playing
different social roles. These roles are acquired through social
interaction and they are not externally imposed or determined by
genetic mechanisms. Rather, the roles {\it emerge} from the
self-organizing dynamics of the complex system. It is here
important to note that the initial differences in number of links
among the agents do not determine the final role of each agent,
since each agent changes temporarily its role in the stochastic
dynamics until a final stationary state is reached. Moreover, we
have checked that taking as initial condition a random network in
which each agent has the same number of links, the system dynamics
leads to the same emergent role differentiation.

Our results indicate that cooperation is stabilized by a
hierarchical self-organized structure, so that the formation of
cohesive clusters of Cooperators with exclusion of free riders is
not needed for persistent global cooperation. Allowing for some
local partner selection, this self-organization leads from a
random network to a final social network with the topology of a
Small World network, giving an example of how the small world
connectivity, in which clustering is larger than in a random
network, can be dynamically achieved.

Our study reveals that if Defectors have the ability to choose
partners, breaking interactions with other Defectors, the number
of Cooperators increases, but the average payoff of Cooperators is
less than that of greedy agents. The dynamics naturally generates
{\it leaders} -individuals getting a large payoff who are imitated
by a considerable fraction of the population-, {\it conformists}
-unsatisfied cooperative agents that keep cooperating-, and {\it
exploiters} -Defectors with a payoff larger than the average one
obtained by Cooperators. The most prominent role is the one of
leader, a Cooperator which not only sustains cooperation, but, it
also drives the whole system towards more cooperation. Defectors
are found to remain in a stable situation whenever they {\em
exploit} Cooperators. The formation of a social hierarchy in the
population is the source of possible unstable behavior. It
promotes the occurrence of social crisis that can affect a large
fraction of the population. These crisis take the form of global
cascades that might be easily triggered by the spontaneous change
of the action of a highly connected agent. This result identifies
the importance of highly-connected agents that, as illustrated by
the imitation network, play a leadership role in the collective
dynamics of the system. Such sensitivity of the network stability
to local special perturbations provides an interesting feature of
globalization: the group of agents organizes itself in a state
where an exogenous or stochastic perturbation may produce drastic
changes, at distance, in a finite time.

Even though our model shows many interesting aspects, the strategy
and neighborhood update rules represent an extreme and
conservative choice of rules. Thus, it would be important to
investigate how the modification of the model rules affect the
results reported here. Some of the points that we consider that
merit further research in connection with our {\it strategy
update} and {\it network update} rules are indicated in the
remainder.

Starting with our strategy update rule, we note that an important
assumption made in our model is that the satisfaction of an agent
is determined by comparison of absolute payoffs. Due to the
evolution of the social interaction this implies that two
neighbors might have different payoff just because they have a
different number of neighbors. On the one hand this incorporates
the idea of the importance of being highly connected. On the other
hand, the possibility of using a comparison based on the relative
payoff per neighbor would imply that each agent knows how many
neighbors has each of its neighbors. Additionally, our strategy
update rule is an extreme ``copy best" rule, while alternatives of
probabilistic imitation of better role models (Schlag and Pollock,
1999), or probabilistic selection of neighbors strategy (Nowak and
May, 1993) could also be considered.

Turning now to the network update rule, we note that in our model
only links between two Defectors can be possibly broken. In a
stationary state most Cooperators are unsatisfied because they
imitate an agent with a larger payoff. In our model this
unsatisfactory situation does not induce an action to improve
their payoff beyond the imitation of a neighbor in the social
network. One way to test the implications of this hypothesis would
be modifying the network update rule, letting the C-D link also to
be broken, for example letting Cooperators have some probability
to change a D-neighbor by a random agent. It is clear that in this
setting cooperation would be further enhanced, and it could also
limit the occurrence of large social crisis. Thus, our rule for
neighborhood update assures that particularly the highly
successful Defectors won't be abandoned by the Cooperators whose
exploitation has made them so successful. Hence, we make Defection
 an attractive role model because exploited agents can not
simply leave their exploiter, e.g., due to costs of relationship
change, or some (semi-)rational calculation of the expected
benefit of creating new ties.
A related point is that one could question why not all D-D links
have the possibility of being broken and not only those that
involve an unsatisfied agent. We have in fact considered that only
agents having at least one neighbor with higher payoff, are
``forced'' to do something to improve their payoff. Thus, we
assume some kind of ``aversion to change'': do something only when
you are not satisfied, otherwise do nothing. This rule can be
justified invoking some cost implicit in the change of a social
relation, and it allows, in principle, to keep the interaction
between two Defectors as long as they are satisfied.

There is another point which deserves further investigation and
which relates both to the strategy and network update rules: Our
choice of strategy update is based on the aggregate payoff, while
our justification of the network adaptation rule is based on
comparison of individual actions. This a priori involves two
different mechanisms and it is clear that our choice is one among
several possible. We have already mentioned a strategy update
\citep{Nowak93} based on probabilistic selection among the
neighbors strategy, weighted by the aggregate payoff.
\citet{Cohen01} compares different such mechanisms on fixed
networks. On the other hand, fewer results are known on network
update mechanisms. In the economics literature a common choice is
that a link among two economic agents is accepted if {\em both}
agents improve their payoff \citep{Jackson99,Bala00}. It would be
interesting to test new strategy and network adaptations involving
a more sophisticated evaluation of the strategy of the opponent
and its aggregate payoff. One possible direction is to contemplate
a more complex strategy space, involving strategies dependent on
the past encounters. Work in this direction with fixed social
network was initiated by \citet{Lindgren94}. Another interesting
direction is to consider non-equivalent agents, so that they may
differ in their attractiveness as exchange partners.
\citet{Flache01} has incorporated this situation in a model that
combines the agents' decisions about cooperation with the
decisions about selection of new partners. They also obtain a
process of social differentiation sustaining cooperation. It is
unclear how much of the process of social differentiation
originates in the unequally attractive agents or on the simpler
mechanisms contained in our model.

Beyond our update rules, the addition of random perturbations in
strategy and network is also a very relevant feature to be
explored in the future. These perturbations may originate from
errors in imitation or payoff determination, for example. Here we
have just shown that strategy perturbations may cause large social
crisis, specially if a well connected agent makes an error.

We finally remark that many contexts in human societies follow the
dynamic scheme considered here -- new collaborations are
frequently formed, while other long lasting partnerships die out.
From scientific collaboration to sports teams --political parties
not to be forgotten--, our study offers an example of simple
mechanism by which leadership, and also other social roles, might
appear and consolidate.

\acknowledgements
VME and MSM acknowledge financial support from MCyT (Spain)
through projects CONOCE; MGZ thanks financial support from
FOMEC-UBA and CONICET (Argentina).


\end{document}